\documentclass{kluwer}    
\usepackage{epsfig}

\begin{document}                                                                                   
\begin{article}
\begin{opening}         
\title{Probing the Circular Polarization of Relativistic Jets on
VLBI Scales} 
\author{Daniel C. \surname{Homan}}  
\institute{National Radio Astronomy Observatory\footnote{The 
National Radio Astronomy Observatory is a 
facility of the National Science Foundation operated under 
cooperative agreement by Associated Universities, Inc.}
\\Charlottesville, VA,22903, U.S.A.}
\author{John F. C. \surname{Wardle}}  
\institute{Brandeis University\\Waltham, MA, 02434, U.S.A.}
\runningauthor{Homan \& Wardle}
\runningtitle{Probing Circular Polarization on VLBI Scales}

\begin{abstract}
High resolution
studies of circular polarization allow us see where it arises
in a jet, study its local fractional level and spectrum, and compare
these results to local measures of linear polarization and Faraday
rotation.  Here we not only review past results from  Very Long
Baseline Array (VLBA) circular polarization studies, but we also
present preliminary new results on two quasars.  In the core of 
PKS 0607$-$157, we find strong circular 
polarization at 8 GHz and much weaker levels at 15 GHz.  Combined
with the linear polarization data, we favor a simple model where the circular
is produced by Faraday conversion driven by a small amount of Faraday
rotation.
In the core of 3C\,345, we find strong circular polarization at 15 GHz
in a component with distinct linear polarization.  This core component
is optically thick at 8 GHz, where we detect no circular polarization. 
With opposite trends in frequency for PKS 0607$-$157 and 3C\,345, it
seems clear that local conditions in a jet can have a strong effect 
on circular polarization and need to be taken into account when
studying inhomogeneous objects with multi-frequency observations.
\end{abstract}


\end{opening}           

\section{Introduction}

Low resolution measurements made during the 1970s and early 1980s
showed circular polarization (CP) to be only a very small fraction of the
integrated synchrotron emission from extra-galactic radio jets.  
In reviewing high accuracy measurements from this
era, \inlinecite{WdP83} found no reliable measurements of CP greater than
$0.5$\% and considered $0.1$\% to be strongly polarized.
Computational jet models by \inlinecite{Jones88} showed that {\em local} levels
of circular polarization could exceed $0.5$\% and that CP was
strongest deep in the jet near the optically thick core where
$\tau\sim 1$.  Jones also found that Faraday conversion of linear to 
circular polarization was the dominant process for producing CP in
his simulated jets.  In the concluding remarks to his 1988 paper,
Jones emphasized the importance of high resolution studies of CP
using the National Radio Astronomy Observatory's planned Very Long 
Baseline Array (VLBA).

In principle, Very Long Baseline Interferometry (VLBI) studies of 
circular polarization allow us to 
see where circular polarization arises in a jet, thereby allowing us
to find its local fractional level, measure its spectrum, and combine
this data with local values for the linear polarization, Faraday
rotation, and optical depth.  Our initial detections of circular
polarization with the VLBA were in the radio galaxy, 3C\,84, and
the high powered quasar, 3C\,279 \cite{HWOR98,WHOR98}.  In the case of
3C\,279 we were able to measure a number of the constraints listed
above and concluded the CP resulted from the Faraday conversion
process, implying a low cutoff in the particle energy distribution
($\gamma_{min} < 20$) and therefore suggesting a predominantly 
$e^+e^-$ plasma for the jet on kinetic luminosity grounds \cite{WHOR98}.
Recent radiative transfer simulations by \inlinecite{RB02} and
\inlinecite{BF02} (see also their contributions to these proceedings),
suggest that our derived cutoff in the particle energy distribution
may not be a strict upper limit, relaxing the need for an $e^+e^-$
plasma.

Ultimately, our ability to extract physical insights about the 
nature of jets from CP observations depends on our ability to
determine the dominant mechanism for circular polarization 
production.  In \inlinecite{WH01}, we reviewed a number of mechanisms 
for producing net circular polarization in radio jets and discussed
some of the difficulties in interpreting circular polarization
observations.  On theoretical grounds, e.g. \inlinecite{Jones88}, the
Faraday conversion process should dominate any contribution from the
intrinsic CP of synchrotron radiation; however, only in the case of 
3C\,279 \cite{WHOR98} is there supporting observational evidence that
Faraday conversion is indeed the dominant process.  The spectral
evidence from other objects presents a confusing picture which could be
consistent with either intrinsic CP or Faraday conversion  
\cite{BFB99,SM99,MKRJ00,Fend00,BBFM01,Fend02}.  

The difficulty in finding a clear observational signature of Faraday 
conversion (or of intrinsic CP for that matter) may be due in part
to the confusion from other parts of the jet in these integrated 
measurements; however, even in the case of the intra-day variable
source PKS 1519$-$273 \cite{MKRJ00}, where interstellar scintillation 
gives an effective resolution of tens of micro-arcseconds, the
observed spectrum does not clearly reveal the mechanism.  A
significant confounding
factor, both in these intra-day variable observations and in our high
resolution VLBA observations, is the likely coincidence of a
circularly polarized signal with the inhomogeneous radio core.  As
discussed in \inlinecite{WH01}, depending on the assumptions one
wishes to make, an inhomogeneous radio core can have a wide range of
spectral index values for circular polarization, irrespective of production
mechanism.

In this article we discuss the study of circular polarization in jets
using VLBI.  In section \ref{s:review} we briefly review some key
results and issues from our earlier works.  In section \ref{s:new} we present new,
multi-frequency VLBA images on two quasars: PKS 0607$-$157 and 3C\,345, and
we evaluate the extent to which our interpretation of these results is
confounded by the inhomogeneous cores in which the circular
polarization is detected.  Finally, we conclude by discussing
goals and prospects for future circular polarization studies with the 
VLBA.

\section{What have we learned so far?}
\label{s:review}

Our circular polarization observations for the jet of 3C\,279
(discussed above) were part of a multi-epoch VLBA monitoring campaign
to study rapid changes in total intensity and polarization of
the most highly active blazars.  The program spanned 6 epochs during
1996 with observations at approximately 2 month intervals.  Five of these epochs
were suitable for circular polarization detection at 15 GHz, and these
epochs were re-analyzed to search for circular polarization, yielding
our results on 3C\,279 as well as multi-epoch detections on 3C\,84,
PKS 0528$+$134, and 3C\,273.  All of these results, as well as a
detailed discussion of our techniques were presented in
\inlinecite{HW99}.  

From this study, we learned that {\em local} levels of circular
polarization could be quite strong in the jets, and we detected 
fractional CP ranging from $0.3-1$\%.  In the case of 3C\,273, the
appearance of CP in the jet core coincided with the appearance of
small amounts of linear polarization ($m_c\approx m_L\approx 0.5$\%)
and a change in opacity in the core from very opaque to more optically
thin as a new jet component began to emerge.  In our final epoch, the
circular polarization spanned the core region of 3C\,273 but appeared most 
strongly associated with the newly emerging component.  Interestingly,
the sign of the circular polarization remained consistent in each
source across all of our epochs; three of these sources were
undergoing strong core outbursts at this time which varied their total 
intensity and linear polarization.  To us, this suggested a magnetic
field structure, responsible for setting the sign of
the circular polarization, which is persistent on a timescales 
at least as long as an individual outburst event.

In \inlinecite{HAW01} we presented results of 5 GHz VLBA circular
polarization observations of 40 sources.  Of these sources, we
detected circular polarization in 11 objects at levels of 
$0.14-0.46$\% local fractional circular polarization.  Our
results were very similar to the integrated measurements of 
\inlinecite{RNS00} who studied a different sample of objects at 5 GHz
with the Australian Compact Telescope Array (ATCA).  Other than a
strong tendency for levels of linear polarization to exceed circular, 
both studies found apparently no relation between observed linear and 
circular polarization in either the integrated \cite{RNS00} or
parsec-scale core \cite{HAW01} measurements.  
One might naturally expect a correlation between linear and circular
polarization as both intrinsic circular polarization and Faraday
conversion require high degrees of field order to produce net CP.
Faraday conversion, in particular, is expected to have a strong 
relationship with linear polarization as it is the linear polarization
that is converted to circular \cite{WH01}.  Of course, it is
possible to destroy any correlation between linear and circular
polarization through either some kind of Faraday depolarization 
or by simply having the circular polarization produced in a region 
much smaller than even our VLBI beam.

\subsection{Sign Consistency of CP}

\inlinecite{Kom84} found that, despite reasonably large fractional
variability in circularly polarized flux, changes in sign of the CP
for a particular source were rare.  Over their $\sim5$ year observing
window, only 2 out of 14 variable CP detected sources showed clear
changes in the sign of their circular polarization.  As described
above, our VLBA results presented in \inlinecite{HW99} also show sign
consistency over a shorter observing window when three of our four
detected objects were undergoing large internal changes resulting 
from developing outbursts.  In \inlinecite{HAW01}, we investigated this
issue further by comparing all recently reported CP measurements to
those made $\sim 20$ years ago.  We found that 5 out of 6 sources 
that were detected both today and in the past had the same sign 
of circular polarization.  While the statistics were only suggestive,
we postulated the existence of a long timescale, persistent magnetic 
field structure, perhaps tied to the black hole/accretion disk system, 
which sets the sign of circular polarization in a particular source
\cite{HW99,HAW01}.

Results presented at this meeting have cast some doubt on this idea
of long timescale CP sign consistency as a general property of 
circularly polarized sources (e.g. H. Aller, these proceedings). 
Two of three of our recently reduced VLBA results also show opposite
signs of CP from those detected historically (Homan and Wardle, in prep).
Upon reflection, we believe there are serious flaws in our comparison
of CP signs from recent results to the historical record as compiled 
by \inlinecite{WdP83} and \inlinecite{Kom84}.  Frequency of
observation is often different between the historical and recent 
results.  In the case of 3C\,84, a predominantly positive sign for
the circular polarization at 15 GHz did not match the historically 
detected negative CP at 5 GHz and below; however, our recent VLBA 
results on 3C\,84 (in prep) show, in addition to positive CP at 15 GHz 
and 22 GHz, clear evidence for predominantly negative CP at 5 GHz.
An additional concern is the use of what are often single epoch
detections to infer the ``preferred sign'' of CP today for comparison
to historical measurements.  
Perhaps the best evidence for long term sign consistency in CP is
for Sgr A* where \inlinecite{BFSB02} used archive observations to show
that the circular polarization maintained the same sign over 18 years

Given that stochastic effects may play a significant role in CP
production in at least some sources, the best approach may be long
term, frequent monitoring campaigns to establish the extent to
which individual sources have a preferred sign of CP and on what
timescale this preferred sign persists.  This issue touches upon 
the connection between the BH/accretion disk system and the magnetic
field structures observed in jets.  Ultimately we would like to know
if the BH/accretion disk imprints a signature in the circular
polarization we observe from jets.  If such a signature exists, does
it change with jet radius (perhaps indicating a reduction in magnetic
flux or a change in pitch angle in any helical field) or even with
time (perhaps indicating a fundamental change in the BH/accretion disk 
system)?

\section{New Results}
\label{s:new}

Here we present dual-frequency, VLBA full polarization images 
of the quasars PKS 0607$-$157 ($z=0.324$) and 3C\,345 ($z=0.595$).  
These sources were observed as part of a small survey for 
circular polarization of 11 powerful AGN at 8 and 15 GHz 
with the VLBA (Homan and Wardle, in preparation).  
The sources were observed on January 3, 1998.

\subsection{PKS 0607$-$157}

Figure 1 displays a total intensity image of the milli-arcsecond
jet of PKS 0607$-$157 at 8 GHz (top frame) and full polarization 
images of the core region at 8 GHz (middle frame) and 15 GHz 
(bottom frame).  The full polarization images of the core region
shows three distinct features in total intensity and linear 
polarization: (1) the radio core, (2) a closely 
separated feature $\simeq 0.8$ milli-arcseconds (mas) from the
core at a position angle of $50$ degrees, and (3) a
well separated feature at $\simeq 2.5$ mas at
a position angle of $\simeq 90$ degrees.  

\begin{figure}
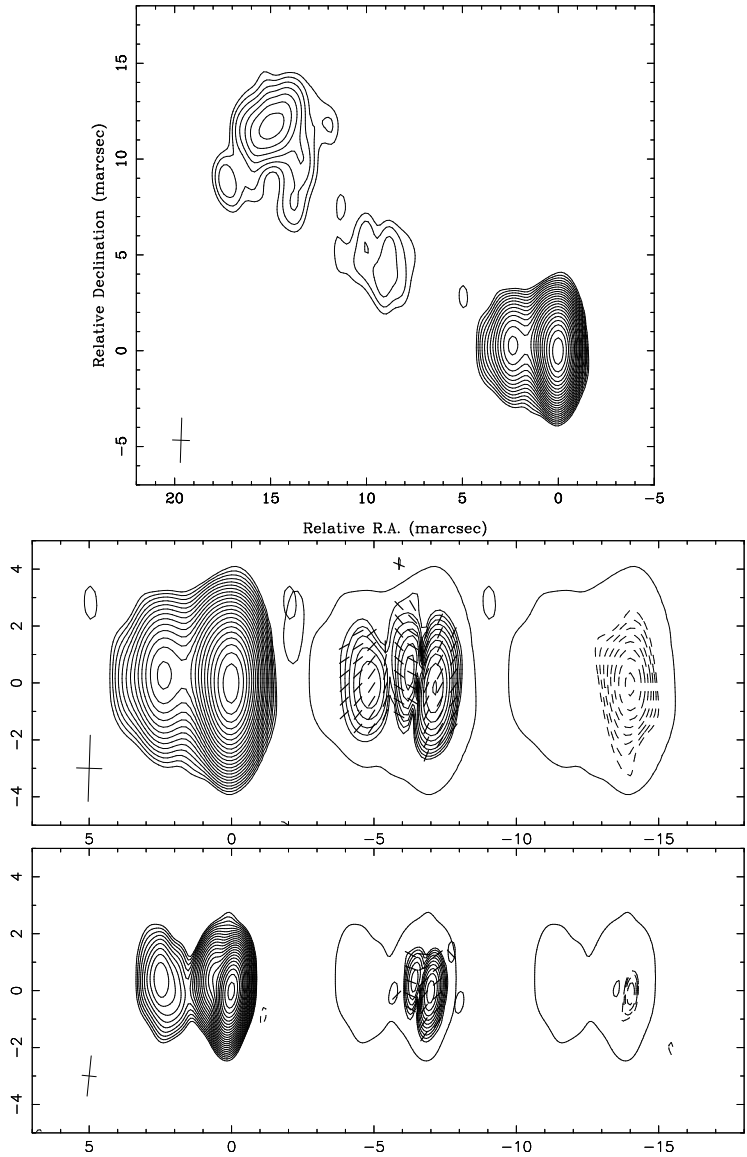

\begin{center}
\epsfig{file=f1a.ps,height=3.0in,angle=270}
\epsfig{file=f1b.ps,height=3.85in,angle=270}
\epsfig{file=f1c.ps,height=3.85in,angle=270}
\end{center}
\caption{\label{f:j0609}
{\em Top frame:} Total intensity image of the milli-arcsecond jet
of PKS 0607$-$157 at 8 GHz.  
{\em Middle and Bottom frames:}  Full 
polarization images of the core region at 8 GHz and 15 GHz
respectively.  The left-hand image is total intensity 
contours.  The middle image is linear polarization intensity 
contours with tick marks representing the electric field vector
direction.  The right image is circular polarization contours 
(dashed=negative).  Both linear and circular polarization images
are surrounded by the lowest contour from the total intensity 
image to show registration.
{\em All frames:} All contours begin at 2 mJy/beam and
increase in steps of $\times\sqrt{2}$.   The FWHM dimensions of the
restoring beam are given by the cross-figure in the lower left-hand
corner of the images. 
}
\end{figure}

For the purposes of this discussion, we will focus only on the 
core and its relation to the closely separated jet feature at $0.8$ mas.  
Preliminary model-fitting shows the core to be optically thick with an
inverted spectral index of $\alpha=0.5$ ($S\propto\nu^{+\alpha}$).
The core has fractional linear polarization at the $2.3$\% level with 
no significant difference between 8 and 15 GHz.  The polarization
angle; however, does show significant rotation, $\Delta\chi_{8-15} =
+16\pm3$ degrees.  Assuming simple  $\lambda^2$ Faraday rotation is
the cause, we calculate a de-rotated position angle of $-48$ degrees 
which suggests a longitudinal magnetic field order pointing within 8 
degrees of the closely spaced jet feature at $0.8$ mas.  Note that
this rotation appears spatially associated with the core, as we find
no detectable rotation in the polarization vectors of the $0.8$ mas
component: $\Delta\chi_{8-15} = -3\pm3$ degrees.

We find the circular polarization to be associated 
with the optically thick radio core.  At 8 GHz the core has 
$-0.65\pm0.07$\% circular polarization, and at 15 GHz we
measure $-0.23\pm0.12$\%.  We also
observed the source a year earlier at 5 GHz, where we found
$-0.18\pm0.05$\% circular polarization on the core \cite{HAW01};
however, here we focus on our simultaneous measurements at 8 and 15 GHz.
Unfortunately, given the much weaker nature of the optically thin
($\alpha=-0.5$) feature at $0.8$ mas, we cannot place useful 
constraints on its fractional circular polarization.

It is unlikely that the very strong CP we observe at 8 GHz is 
intrinsic to the emitted synchrotron radiation.  
We consider a simple model with longitudinal field order along the 
jet axis and an additional random component of magnetic field (which 
may be shocked but not strongly enough to overcome the net
longitudinal field order).  In this model, the longitudinal field is 
responsible for both the linear and circular polarization we observe, 
and we estimate that a very high fraction of the total field energy in 
the jet, $\gtrsim 40-50$\%, must be in the form of magnetic flux to
explain the observed circular polarization by the intrinsic mechanism 
alone.  Assuming a conically expanding jet, the resulting square-law 
dependence of magnetic flux on radius makes it 
difficult to generate less circular polarization at 15 GHz than
at 8 GHz via the intrinsic mechanism; however, we note that the core 
is inhomogeneous and it is possible that the local jet conditions 
(i.e. density, field strength) at the 15 GHz optical surface do not 
follow a simple extrapolation from the 8 GHz core.   

Faraday conversion of linear to circular polarization can 
reproduce the observed levels of circular polarization at both 8
and 15 GHz without any requirement for large quantities of magnetic
flux in the jet.  Here we assume the Faraday conversion is driven by 
a small amount of Faraday rotation.  Indeed the very rotation we
observe in the linear polarization of the core between 8 and 15 GHz,
if interpreted as resulting from a small internal Faraday depth, is 
of approximately the right amount (and is the correct sign) to drive 
the conversion process and produce the CP we observe.  The 
smaller degree of rotation expected at 15 GHz also explains the 
reduced levels of CP measured at that frequency.  

In this model, the low degree of observed linear polarization
($2.3$\%) requires a high efficiency for the Faraday conversion 
process to produce
$0.6-0.7$\% CP at 8 GHz.  This requirement for a high efficiency
in the Faraday conversion process is paired with an apparently 
small amount of Faraday rotation ($\sim 20^\circ$ by 8 GHz).  
Of course, having only two frequencies restricts our ability to 
robustly say that any internal Faraday rotation is indeed small, but
there are a number of lines of supporting evidence: (1) a low degree
of apparent linear de-polarization ($2.5\pm0.2$\% linear polarization 
at 15 GHz and $2.2\pm0.2$\% at 8 GHz), (2) the apparent alignment of
the de-rotated magnetic field with the position angle of the first jet 
feature, and 
(3) the consistent prediction for the reduced degree of CP at 15 GHz.  Highly
efficient Faraday conversion demands a small $\gamma_{min}$, the
low-energy cutoff in the particle power-law spectrum; however a small
value of $\gamma_{min}$ produces a large amount of Faraday rotation,
which we argue is not observed.  To suppress the Faraday rotation, we
need some combination of (1) many field reversals along the line of sight
and/or (2) a significant population of $e^+e^-$ pairs which produce
no Faraday rotation. 

One alternative to this small Faraday rotation depth model is the high 
rotation depth model of \inlinecite{BF02} and
\inlinecite{RB02}, which can
simultaneously depolarize the linear polarization and produce 
significant quantities of circular polarization.  While we see no
evidence for strong Faraday rotation or depolarization between 8 and 
15 GHz, the inhomogeneous nature of the core region may allow a fixed 
amount of depolarization with frequency and an apparently small degree
of rotation between the frequencies.

\subsection{3C\,345}

\begin{figure}
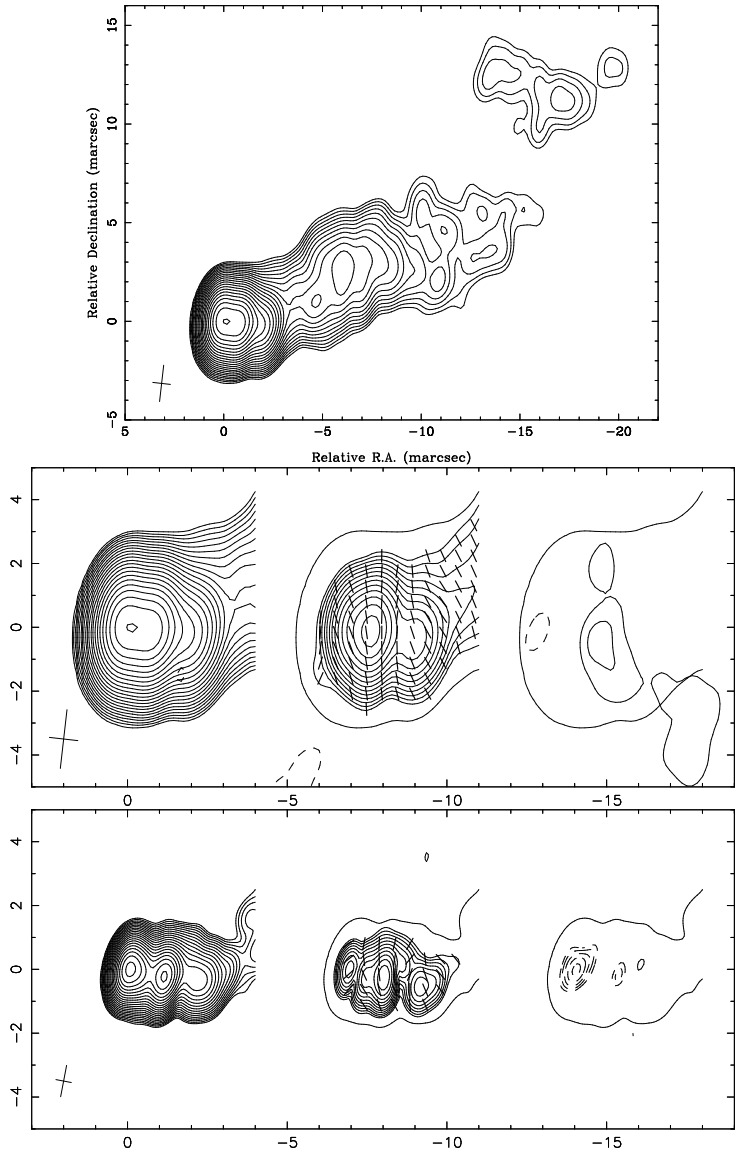

\begin{center}
\epsfig{file=f2a.ps,height=3.0in,angle=270}
\epsfig{file=f2b.ps,height=3.8in,angle=270}
\epsfig{file=f2c.ps,height=3.8in,angle=270}
\end{center}
\caption{\label{f:j1642}
{\em Top frame:} Total intensity image of the milli-arcsecond jet
of 3C\,345 at 8 GHz.  
{\em Middle and Bottom frames:}  Full 
polarization images of the core region at 8 GHz and 15 GHz
respectively.  The left-hand image is total intensity 
contours.  The middle image is linear polarization intensity 
contours with tick marks representing the electric field vector
direction.  The right image is circular polarization contours 
(dashed=negative).  Both linear and circular polarization images
are surrounded by the lowest contour from the total intensity 
image to show registration.
{\em All frames:} All contours begin at 2 mJy/beam and
increase in steps of $\times\sqrt{2}$.   The FWHM dimensions of the
restoring beam are given by the cross-figure in the lower left-hand
corner of the images. 
}
\end{figure}

Figure 2 displays a total intensity image of the milli-arcsecond
jet of 3C\,345 at 8 GHz (top frame) and full polarization 
images of the core region at 8 GHz (middle frame) and 15 GHz 
(bottom frame).  We only detect circular polarization at 15 GHz,
where we see the core component with $-0.38\pm0.12$\% CP.  This core
component also has a distinct linear polarization which is clearly
visible at 15 GHz.  At 8 GHz, this component must be quite optically
thick as we do not see its contribution in either linear or circular
polarization.   

Without multi-frequency constraints, it is difficult to analyze the 
circular polarization in detail; however, we note that the lack
of a detectable signal at the location of the 8 GHz optical surface 
indicates that local conditions in the jet are important to
the production of circular polarization.  One might be able to
construct simple scaling laws to explain the disappearance of 
circular polarization at lower frequency, but PKS 0607$-$157, which
we discuss above, shows the opposite trend. 

\section{Conclusions}

VLBI studies of circular polarization can provide local measures of
both linear and circular polarization at multiple frequencies,
which, in principle, allow detailed analysis of the emission mechanism 
and the physics constrained by it.  
In the case of 0607$-$157, our preliminary analysis favors 
Faraday conversion driven by a small amount of Faraday rotation;
however, we are unable to exclude at least two alternative models due, 
in part, to the inhomogeneous nature of the emission region.  We also
present new results on 3C\,345 which shows strong CP at 15 GHz and no 
detectable CP at 8 GHz.  Coupled with our results on PKS 0607$-$157,
which has the reversed trend with distinctly stronger CP at 8 GHz, it 
seems clear that local conditions in the jet can have a strong effect 
on circular polarization and need to be taken into account when
studying inhomogeneous objects with multi-frequency observations.

Future detailed VLBI studies of circular polarization would greatly 
benefit from expanded frequency coverage, particularly coverage within 
bands by using widely separated IF channels.  Of key importance will
be our ability to tie down
not only the circular polarization spectrum but also the observed
Faraday rotation.  Such detailed studies require increased sensitivity
and a reliable list of calibrator sources, so that an observer can
obtain simultaneous, deep observations at each frequency.  Enhanced
sensitivity and improved calibration will also increase our
ability to detect (or at least usefully limit) circular polarization 
in optically thin jet features.  Such features would be much easier to 
model and study than inhomogeneous radio cores; unfortunately, with a 
handful of exceptions, discrete optically thin jet features are too weak
to be studied by our current techniques.

\acknowledgements
This work has been supported by the National Radio Astronomy
Observatory and by NSF grant AST 99-00723.


\def\ref@jnl#1{{\it#1}}

\def\aj{\ref@jnl{AJ}}                   
\def\araa{\ref@jnl{ARA\&A}}             
\def\apj{\ref@jnl{ApJ}}                 
\def\apjl{\ref@jnl{ApJ}}                
\def\apjs{\ref@jnl{ApJS}}               
\def\aap{\ref@jnl{A\&A}}                
\def\mnras{\ref@jnl{MNRAS}}             
\def\nat{\ref@jnl{Nature}}              

\end{article}
\end{document}